\documentclass[a4paper,10pt]{article}
\usepackage[utf8]{inputenc}
\usepackage{amssymb,amsmath}
\usepackage{authblk}
\usepackage[left=0.5in, right=0.5in, top=1in, bottom=1in]{geometry}
\usepackage{hyperref}
\usepackage{graphicx}
\usepackage{subfig}

\title{A proof  to Biswas-Mitra-Bhattacharyya conjecture
for ideal quantum
gas trapped under generic power law potential $U=\sum_{i=1} ^d c_i |\frac{x_i}{a_i}|^{n_i}$ in $d$ dimension}

\author{Mir Mehedi Faruk$^{a,b}$, Md. Muktadir Rahman$^{a}$\\
Department of Theoretical Physics, University of Dhaka, Dhaka-1000$^a$\\
Theoretical Physics, Blackett Laboratory, Imperial College, London SW7 2AZ, United Kingdom$^b$\\
\href{mailto:me@somewhere.com}{Email: muturza3.1416@gmail.com, mmf15@ic.ac.uk, mir.mehedi.faruk@cern.ch, remon.pp@gmail.com} 
 }
 
\begin{document}
\maketitle
 
 \begin{abstract}
 The well known relation for ideal 
 classical gas $\Delta \epsilon^2=kT^2 C_V$ which 
 does not remain valid for quantum system is revisited. A new connection
 is established between
 energy fluctuation and specific heat for quantum gases, valid in the classical limit and the degenerate quantum regime
  as well. Most importantly the proposed
  Biswas-Mitra-Bhattacharyya (BMB)
   conjecture 
 (Biswas $et.$ $al.$, J. Stat. Mech. P03013, 2015.)\cite{s}
 relating hump in energy fluctuation and discontinuity of specific heat 
 is proved and precised in this manuscript.
 \end{abstract}

\section{Introduction}
An increasing attraction towards the 
subject of quantum gases is observed,
after it was possible to create BEC in
magnetically trapped alkali gases\cite{Bradley,anderson,davis} as well as experimental confirmation  
of Fermi degeneracy\cite{DeMarco}.
Different theoretical and experimental
studies analysing
the effects of
temperature dependence of energy and specific heat of ultracold Fermi
gases\cite{DeMarco,Kinast,Biswas}, momentum distribution for harmonically trapped quantum gas\cite{Truscott,Luo},
 temperature dependence of the
chemical potential\cite{Lou1}, critical number of particles
for the collapse of attractively interacting Bose gas\cite{Biswas1}, Casimir effect\cite{casimir,marek1},
equivalence of Bose and Fermi system\cite{turza,lee}, q deformed systems\cite{q1,q2}
have already been reported. 
Although a lot of theoretical 
studies\cite{tf,pathria,huang,ziff,sala,yan,yan1,yan2}
are done on quantum gases trapped under generic power law potential, none of these contained
detailed discussion on energy fluctuation of trapped quantum gases,
until the recent paper of Biswas et. al. \cite{s}.
In the case of
ideal classical gas,
specific heat $C_V$ is regarded as energy fluctuation $\Delta \epsilon^2$ as
the  $\Delta \epsilon^2$
is related to  $C_V$ as, $\Delta \epsilon^2= kT^2 C_V$. But this relation becomes
invalid for both types of quantum gases (Bose and Fermi) in the quantum degenerate regime for
free quantum gases\cite{s}. 
 Biswas $et.$ $al.$\cite{s} have conjectured (BMB conjecture) 
 a relation between the discontinuity of $C_V$ and energy fluctuation. 
 According to the BMB conjecture,
the appearance of a hump in $\frac{\Delta \epsilon^2}{ kT^2}$  over its classical limit may
indicate a discontinuity of $C_V$. They have shown this to be true for free and harmonically
trapped Bose gases\cite{s}. They have also mentioned without proving that the 
the inverse of this statement may not
always be true. But this conjecture is yet to be proven for any  quantum system 
trapped under generic power
law potential in arbitrary 
dimension and is an open problem. In this manuscript, we have proved and precised 
BMB conjecture for ideal quantum gases trapped under
generic power law potential, $U=\sum_{i=1} ^d c_i |\frac{x_i}{a_i}|^{n_i}$
in $d$ dimension. Thus, in principle one can reconstruct the
results  Shyamal $et.$ $al.$, choosing all $n_i=2$ for harmonically  trapped system and all 
$n_i=\infty$ for free systems.
Beside this, a relation is established between $C_V$ and
${\Delta \epsilon^2}$ which is valid  not only  in 
the classical limit but  in the quantum limit as well.\\\\
The manuscript is organized in the following way.
In section 2,
we have determined  the grand potential in an unified
way for both types of quantum gases, from 
which we are able to calculate the quantities such as $C_V$
and $\Delta \epsilon^2$.
In the next section we elaborate two 
theorems which eventually guide us to prove the conjecture. 
Results are discussed in section 4 and
the paper is concluded in section 5.

\section{Grand potential, specific heat and energy fluctuations}
For a quantum gas, the average number of particles occupying the $i$-th eigenstate and the grand potential 
, are given by
\begin{eqnarray}
\bar{n}_i &=&\frac{1}{z^{-1}e^{\beta\epsilon_i}-a}
\end{eqnarray}and,
\begin{eqnarray}
q&=&\frac{1}{a}\sum_\epsilon ln(1+az e^{-\beta \epsilon})
\end{eqnarray}
where, $a=1(-1)$ stands for Bose system (Fermi system)
and $z$ is fugacity.
Let us consider an ideal quantum system trapped in a generic power law potential
in $d$ dimensional space with  a single particle Hamiltonian of the form,
\begin{eqnarray}
 \epsilon (p,x_i)= bp^l + \sum_{i=1} ^d c_i |\frac{x_i}{a_i}|^{n_i}
\end{eqnarray}\\
where, $b,$ $l,$ $a_i$, $c_i$, $n_i$  
are all positive constants, $p$ is the momentum 
and $x_i$ is the  $i$ th component of coordinate of a particle.
Here, $c_i$, $a_i$, $n_i$ determines the depth 
and confinement power of
the potential and $l$ being the kinematic parameter and $x_i<a_i$. As $|\frac{x_i}{a_i}|<1$,
the potential term goes to zero
as all $n_i\longrightarrow \infty$.
 Using $l=2$, $b=\frac{1}{2m}$ one can get the energy spectrum  of  the hamiltonian used in 
the literatures \cite{pathria,huang,ziff,sala}.
If one uses $l=1$ and $b=c$ one finds the hamiltonian of massless systems such as photons\cite{huang}.
\\\\
The density of states for such system is \cite{tf,yan2},
\begin{eqnarray}
 \rho(\epsilon)=C(b,V_d')\epsilon^{\chi-1}
\end{eqnarray}
where, $C(b,V_d')$ is a constant depending on effective volume $V_d'$\cite{turza}.
For the detail derivation of density of states see Ref. \cite{yan2}.
Now, replacing the sum by integral we obtain the grand potential,
\begin{eqnarray}
q=q_0+\frac{V_d'}{\lambda'^d}Li_{\chi+1}(\sigma) 
\end{eqnarray}
Note, $Li_{q}(m)$ is known as polylog function in the literature whose
 integral representation   for $Re(m)<1$
is\cite{lee}
\begin{eqnarray}
 &&Li_{q}(m)=\frac{1}{\Gamma(q)}\int_0 ^m [\ln(\frac{m}{\eta})]^{q-1}\frac{d\eta}{1-\eta}, \\
&&q_o=\frac{1}{a}\ln(1+az)
 \end{eqnarray}
 It is  a real valued function if $m\in \mathbb{R}$ and $-\infty<m<1$. Also, the effective volume $V_d'$, 
 effective  thermal wavelength $\lambda_d '$ along with $\chi$ and
 $\sigma$ are defined as,
\begin{eqnarray}
      V_d ' &=& V_d \prod_{i=1} ^d (\frac{kT}{c_i})^{1/n_i}\Gamma(\frac{1}{n_i} + 1)\\
     \lambda'&=& \frac{h b^{\frac{1}{l} }}{\pi ^{\frac{1}{2}} (kT) ^{\frac{1}{l}}} [\frac{d/2+1}{d/l+1}]^{1/d}\\
     \chi&&= \frac{d}{l} + \sum_{i=1} ^d \frac{1}{n_i}\\
   \sigma &&= \left\{
     \begin{array}{lr}
     -z &,  \text{Fermi system}\\
     z &,  \text{Bose system}
     \end{array}
   \right.
 \end{eqnarray}
 The detail idea of effective thermal wavelength
 and effective volume for trapped quantum gases can be found in\cite{yan,yan1, yan2, yan3}.
But note, when $l=2$ and $b=\frac{1}{2m}$ from Eq. (8) 
we get $\lambda_0=\frac{h}{(2\pi mk T)^{1/2}}$, which is the thermal 
wavelength of nonrelativistic  massive fermions as well as  massive bosons.
However, it should be noted that, 
when $l\neq 2$,  $\lambda'$ 
then depends on dimension. With $d=3$ and $d=2$, thermal wavelength of photons (boson) and neutrinos (fermion) are 
respectively\cite{pathria} $\frac{hc}{2\pi^{1/2} kT}$ 
and $\frac{hc}{(2\pi)^{1/2}kT}$ which can be obtained from from Eq. (9) by choosing $b=c$, where $c$ being
the velocity of light.
But one needs to consider the effects of 
antiparticles to calculate the thermodynamic quantities
of ultrarelativistic quantum gas\cite{howard}.
So, one can reproduce the thermal wavelength of  both massive and massless particles 
from the definition of effective thermal wavelength\footnote{For detailed conceptual 
overview of effective thermal wavelength
see Ref. \cite{yan3}}
$\lambda '$  with more general energy
spectrum.  
 It is also seen that 
the effective volume\footnote{Effective volume is also referred as pseudovolume, 
for detailed conceptual overview on how equation of state of trapped quantum system can be obtained see Ref.\cite{yan1} } $V_d '$
is a very salient feature of the  trapped system as this allows us to treat the trapped quantum
gases\cite{turza,yan,yan1,yan2} 
to be treated as a free one.   Difference between $V_d'$  and $V_d$
is that, the former depends on temperature and power law exponent while the latter does not.
But as all $n_i\longrightarrow \infty$, $V_d'$ approaches $V_d$.
The great benefit of evaluating $V_d'$ and $\lambda '$ is that they 
enable us to write all of the thermodynamic functions of the trapped quantum system to be expressed
in a compact form similar to those of a free quantum gas\cite{turza,yan}.
 It is well known that the Bose and Fermi functions do represent 
thermodynamics of Bose and Fermi system and
 can be written in terms of Polylogarithmic functions,
\begin{eqnarray}
&& Li_l(z)=g_l(z)= \frac{1}{\Gamma(l)}\int_0 ^\infty dx \frac{x^{l-1}}{z^{-1}e^x-1}\\
 &&- Li_l(-z)=f_l(z)=\frac{1}{\Gamma(l)}\int_0 ^\infty dx \frac{x^{l-1}}{z^{-1}e^x+1}
\end{eqnarray}
Point to note,
as $z\rightarrow 1$, the Bose function $g_\chi(z)$ approaches $\zeta(\chi)$, for $\chi>1$\cite{pathria}.
From the grand potential we can now  determine, 
\begin{eqnarray}
 &&E=-(\frac{\partial q}{\partial \beta})_{z,V_d '}=NkT\chi\frac{Li_{\chi+1}(\sigma)}{Li_{\chi}(\sigma)}\\
  &&C_V=(\frac{\partial U}{\partial T})_{N, V_d'} 
  =Nk[\chi(\chi+1)  \frac{Li_{\chi+1}(\sigma)}{Li_{\chi}(\sigma)} - \chi^2\frac{Li_{\chi}(\sigma)}{Li_{\chi-1}(\sigma)} ]
\end{eqnarray}
In the classical limit of quantum gases, $C_V$ equals to $N k\chi $. Now, the energy fluctuation 
\begin{eqnarray}
 \Delta \epsilon^2&=&\bar{\epsilon^2}-\bar{\epsilon}^2=\sum_i \bar{n}_i \epsilon_i ^2 - (\sum_i \bar{n}_i \epsilon_i)^2
 =\int d\epsilon \rho(\epsilon) \epsilon^2 n(\epsilon)  - (\int d\epsilon \rho(\epsilon)  \epsilon n(\epsilon) )^2\nonumber \\
&=& (kT)^2[\chi(\chi+1)\frac{Li_{\chi+2}(\sigma)}{Li_{\chi}(\sigma)}-\chi^2\frac{Li_{\chi+1} ^2(\sigma)}{Li_{\chi} ^2(\sigma)}]
 \end{eqnarray}
 So, it is clear from Eq. (15) and (16) that, $\Delta \epsilon ^2 \neq kT^2 C_V$.
 But it is valid in  the classical limit as $z\rightarrow 0$,
 Eq. (15) and (16) deptics,  
\begin{equation}
 \Delta \epsilon ^2 = kT^2 C_V = N\chi (kT)^2 
\end{equation} 
But we can establish such a relation valid within the whole temperature range.
Note, 
\begin{eqnarray}
\Delta \epsilon^2=\bar{\epsilon^2}-\bar{\epsilon}^2
&=&(\frac{\partial E}{\partial \beta})_{_{z,V_d'}}
=kT^2(\frac{\partial E}{\partial T})_{_{z,V_d'}}
=kT^2(\frac{\partial E}{\partial T})_{_{N,V_d'}}+
kT^2(\frac{\partial E}{\partial N})_{_{T,V_d'}}(\frac{\partial N}{\partial T})_{_{z,V_d'}}\nonumber \\
&=&kT^2 C_V+
kT(\frac{\partial E}{\partial N})_{_{T,V_d'}}(\frac{\partial E}{\partial \mu})_{_{T,V_d'}}
\end{eqnarray}
where, in the last line we have used
$\frac{1}{T}(\frac{\partial E}{\partial \mu})_{_{T,V_d'}}=(\frac{\partial N}{\partial T})_{_{z,V_d'}}$\cite{pathria}. 
In the high temperature limit the second term of Eq. (18) becomes zero and Eq. (18) coincides with Eq. (17). 
It can be easily  justified  that equation (18) is valid 
not only in the classical limit but also
in the quantum degenerate regime.
\\\\
Point to note that expression of $C_V$ and
$\Delta \epsilon^2$ represents both Bose and Fermi system in a unified approach.
In the case of Fermi system,
\begin{eqnarray}
&&C_V =Nk[\chi(\chi+1)  \frac{f_{\chi+1}(z)}{f_{\chi}(z)} -
 \chi^2\frac{f_{\chi}(z)}{f_{\chi-1}(z)} ]\\
&& \Delta \epsilon^2=(kT)^2[\chi(\chi+1)\frac{f_{\chi+2}(z)}{f_{\chi}(z)}-\chi^2\frac{f_{\chi+1} ^2(z)}{f_{\chi} ^2(z)}]
\end{eqnarray}\\
The above equations coincides exactly with the results of Ref. \cite{s}
appropriate choice of $n_i$ and $d$. Writing the expressions for  Bose  system (per particle),
\begin{eqnarray}
  && C_V =\left\{
     \begin{array}{lr}
        k [\chi(\chi+1)\frac{\nu'}{\lambda '^D}g_{\chi+1}(z)-\chi^2 \frac{g_{\chi}(z)}{g_{\chi-1}(z)}] &,  T>T_c\\
        k \chi (\chi +1)\frac{\nu'}{\lambda '^D} \zeta(\chi+1)&,  T\leq T_c
     \end{array}
   \right.\\
   &&\Delta \epsilon^2 =\left\{
     \begin{array}{lr}
      (kT)^2[\chi(\chi+1)\frac{g_{\chi+2}(z)}{g_{\chi}(z)}-\chi^2\frac{g_{\chi+1} ^2(z)}{g_{\chi} ^2(z)}] &,  T>T_c\\
    (kT)^2[   A_1 (\frac{T}{T_C})^\chi  - A_2 (\frac{T}{T_C})^{2\chi}] &,  T\leq T_c
     \end{array}
   \right.  
\end{eqnarray}       
  Here $T_C$ denotes  critical temperature
and, $A_1$ and $A_2$
are defined as,
\begin{eqnarray}
 A_1&=&\chi(\chi+1)\frac{\zeta(\chi+2)}{\zeta(\chi)}\\
 A_2&=&[\chi\frac{\zeta(\chi+1)}{\zeta(\chi)}]^2
 \end{eqnarray}
 Eq. (22) agrees with the  expressions for free and harmonically trapped Bose system 
 in three dimensional space\cite{s}. 
 It is noteworthy that, for both types
 of trapped quantum gases, $C_V$ approaches its classical value $\chi N K$ \cite{t1}.

\section{Theorems regarding $C_V$ and $\Delta \epsilon ^2$ of Bose gas}
In this section we present  the necessary theorems to prove the conjecture.
As it was shown by Shyamal $et.$ $al.$ that a hump does exist in ${\Delta \epsilon^2}/{kT^2 C_V ^{^{cl}}}$
 in the condensed phase for
harmonically trapped Bose gas,
 we need to find a general criteria to locate a  hump in ${\Delta \epsilon^2}/{kT^2 C_V ^{^{cl}}}$ 
  over its classical limit in arbitrary dimension with any trapping potential. As there is no condensed phase in ideal 
  Fermi gas trapped under potential\cite{t2}, this theorem bears no significance for them. 
   \\\\
\textbf{Theorem 4.1}: {\em Let an ideal Bose gas in an external potential,
$U=\sum_{i=1} ^d c_i |\frac{x_i}{a_i}|^{n_i}$. A hump will exist in the condensed phase  in
${\Delta \epsilon^2}/{kT^2 C_V ^{^{cl}}}$  over the classical limit if,
\begin{eqnarray}
&&\chi < \frac{A_1^2}{4 A_2} =\gamma(\chi) \nonumber\\
\Rightarrow &&\chi\geqslant 2.3 \nonumber
\end{eqnarray}} 
\textbf{Proof}:\\\\
Re-writing Eq. (22) in the condensed phase, 
\begin{eqnarray}
\frac{\Delta \epsilon^2}{(kT)^2}=f(\tau)=A_1 \tau ^\chi -A_2 \tau ^{2\chi} \nonumber\\
\Rightarrow f'(\tau) = \frac{\partial f}{\partial \tau}= A_1 \chi \tau^{\chi-1} - 2 A_2 \chi \tau^{2\chi -1}
\end{eqnarray}
The condition for maximum,
\begin{eqnarray}
&& f'(\tau)\bigg|_{\tau = \tau_0} = 0\nonumber\\
&& \Rightarrow \tau_0^\chi=\frac{A_1}{2 A_2} 
\end{eqnarray}\\
The hump will be over its classical limit if, 
\begin{eqnarray}
&& f(\tau)\bigg|_{\tau = \tau_0} > \chi	\nonumber\\
\Rightarrow && A_1 \tau_0^\chi - A_2 \tau_0^{2 \chi} > \chi \nonumber\\
\Rightarrow && \chi < \gamma(\chi)=\frac{A_1^2}{4 A_2}  
\end{eqnarray}\\
From the Table 1,
one can conclude relation (27) will be maintained
(a hump will exist in the condensed phase over the classical limit) when,
\begin{equation}
\chi\geqslant 2.3 
\end{equation}\\
\begin{table}
\begin{center}
\caption{Listed values of $\gamma(\chi)$, generated using Mathematica (Correct upto four decimal points).}
\begin{tabular}{|c|c|}
\hline
$\chi$ & $\gamma(\chi)$ \\[1ex]
\hline\hline
1.3 & 0.8553 \\
1.4 & 0.9756 \\
1.5 & 1.1022 \\
1.6 & 1.2350 \\
1.7 & 1.3736 \\
1.8 & 1.5181 \\
1.9 & 1.6683 \\
2.0 & 1.8240 \\
2.1 & 1.9853 \\
2.2 & 2.1519 \\
2.3 & 2.3239 \\
2.4 & 2.5010 \\
2.5 & 2.6834 \\
2.6 & 2.8709 \\
\hline
\end{tabular}
\end{center}
\end{table}\\
\textbf{Theorem 4.2}: {\em Let an ideal Bose gas in an external potential,
$U=\sum_{i=1} ^d c_i |\frac{x_i}{a_i}|^{n_i}$, \\\\
(i) $C_V$ will be discontinuous at $T=T_c$
if,
\begin{eqnarray}
\chi=\frac{d}{l}+\sum_{i=1} ^d \frac{1}{n_i}>2 \nonumber
\end{eqnarray}}. \\
(ii) And the difference between the heat capacities at constant volume, at $T=T_c$ as
\begin{eqnarray}
\Delta C_V\mid_{_{T=T_c}}=C_V\mid_{_{T_c^{-}}}- C_V\mid_{_{T_c^{+}}}=Nk\chi^2\frac{\zeta(\chi)}{\zeta(\chi-1)}\nonumber
\end{eqnarray}
\\\\
\textbf{Proof}:\\\\
As $T\rightarrow T_c$, $z\rightarrow 1$ and $\eta\rightarrow 0$, where $\eta=-\ln z$.
For $T\rightarrow T_c ^+$, 
\begin{eqnarray}
C_V (T_C^+) &&=     N k [\chi(\chi+1)\frac{\nu'}{\lambda '^D}g_{\chi+1}(z)|_{_{z\rightarrow 1}}
-\chi^2 \frac{g_{\chi}(z)}{g_{\chi-1}(z)}|_{_{z\rightarrow 1}}] \nonumber \\
&& =N k [\chi(\chi+1)\frac{\nu'}{\lambda '^D}\zeta({\chi+1})- \chi^2 \frac{\zeta({\chi})}{g_{\chi-1}(z)}|_{_{z\rightarrow 1}}]\
\end{eqnarray}
As the denominator of the second term of the right hand side 
contains $g_{\chi-1}(z)$, we can not simply substitute it with zeta function as $z\rightarrow 1$.
So, using the  representation of Bose function by Robinson\cite{r},
\begin{eqnarray}
 g_\chi(e^{-\eta})=\frac{\Gamma(1-\chi)}{\eta ^{1-\chi}} + \sum_{i=0} ^{\infty} \frac{(-1)^i}{i!}\zeta(\chi-i)\eta^i
\end{eqnarray}
we get from the above,
\begin{eqnarray}
 C_V (T_C^+) = N k [\chi(\chi+1)\frac{\nu'}{\lambda '^d}\zeta({\chi+1})-
 \chi^2 \frac{\zeta({\chi})}{\Gamma(2-\chi)} \eta ^{2-\chi}\mid _{\eta\rightarrow 0}]
\end{eqnarray}
On the other hand 
\begin{eqnarray}
 C_V (T_C^-) = N k [\chi(\chi+1)\frac{\nu'}{\lambda '^d}\zeta({\chi+1})
\end{eqnarray}\\
Taking the difference between $C_V(T_C ^+)$ and $C_V(T_C ^-)$, we get,
\begin{eqnarray}
\Delta C_V\mid_{_{T=T_c}}=  \chi^2 \frac{\zeta({\chi})}{\Gamma(2-\chi)} \eta ^{2-\chi}\mid _{\eta\rightarrow 0}
\end{eqnarray} 
Which dictates, $\Delta C_V\mid_{_{T=T_c}}$ 
will be non zero for $\chi>2$. So, $C_V$ will be discontinuous
when $\chi>2$
and thus completing first part of the theorem.\\\\
As, 
$\chi$ should be greater than two for $\Delta C_V$ at $T=T_C$ to be non-zero, we can re-write 
equation (21), by substituting $g_{\chi-1}(z)$ by zeta function.
\begin{eqnarray}
C_V (T_C^+) = N k [\chi(\chi+1)\frac{\nu'}{\lambda '^D}\zeta({\chi+1})- \chi^2 \frac{\zeta({\chi})}{\zeta({\chi-1})}]
\end{eqnarray}
 From Eq. (32) and (34) we can write.
\begin{eqnarray}
\Delta C_V\mid_{_{T=T_c}}=C_V\mid_{_{T_c^{-}}}- C_V\mid_{_{T_c^{+}}}=Nk\chi^2\frac{\zeta(\chi)}{\zeta(\chi-1)}
\end{eqnarray}
Note that, the same result is also derived 
by Yan et. al. for Bose gas trapped in symmetric power law potential\cite{yan2}.  
\\ \\
Now, based on the above two theorems we can come to the  conclusion. We have seen, 
a hump will exist in ${\Delta \epsilon^2}/{kT^2 C_V ^{^{cl}}}$  over the classical limit when $\chi\geqslant 2.3$ and 
$C_V$ will be discontinuous while $\chi>2$. Therefore the existence of hump in ${\Delta \epsilon^2}/{kT^2 C_V ^{^{cl}}}$
over the classical limit automatically ensures discontinuity in $C_V$. But a discontinuity in $C_V$ does not
imply a hump in ${\Delta \epsilon^2}/{kT^2 C_V ^{^{cl}}}$  over the classical limit when the value of $\chi$ is in between,
$2<\chi<2.3$ and thus, proving the conjecture.\\
\section{Results and Discussion}
In this section we discuss energy fluctuation and specific heat in detail 
and check the prediction of the above theorems. We have illustrated in this paper how specific 
heat can differ from the energy fluctuation for the entire range of temperature. It is important to 
note that the difference between
the probabilities of the classical
and the quantum gas arises essentially from the nonzero fugacity of the quantum gas. \\\\
\begin{figure}[h!]
\centering
\includegraphics[ height=10cm, width=11cm]{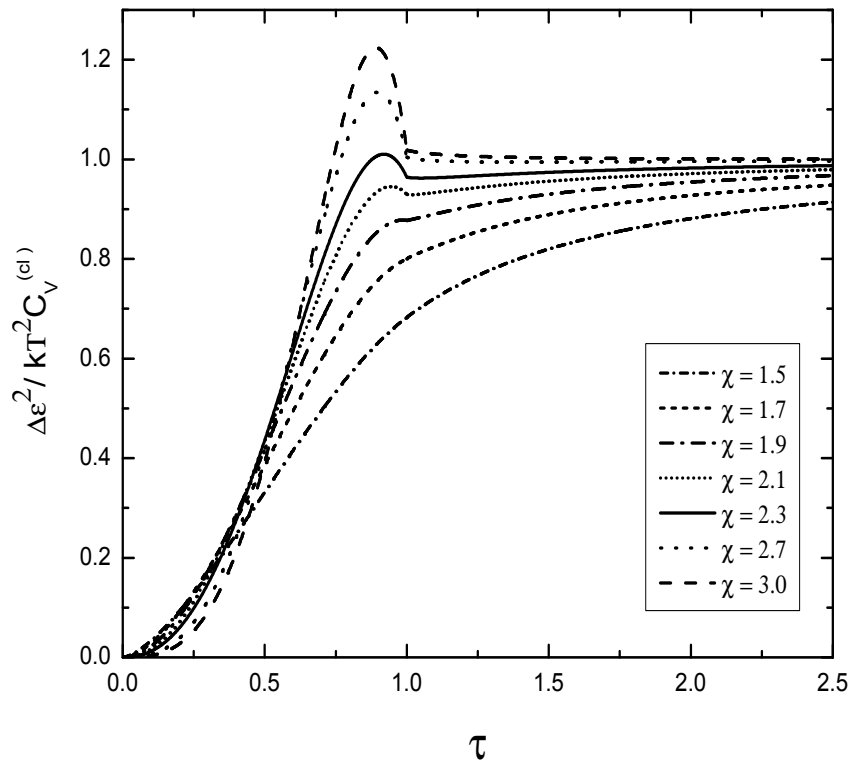}  
\caption{Energy fluctuation ideal trapped Bose gas
as a function of $\tau=\frac{T}{T_C}$, with different power law potentials. }
  \label{fig:boat1}
\end{figure}
\begin{table}[h!]
\begin{center}
\caption{Status of energy
fluctuation and specific heat of Bose system trapped under generic power law potential.}
\begin{tabular}{|c|c|c|}
\hline
Range of $\chi $ & Hump over classical limit in ${\Delta \epsilon^2}/{kT^2 C_V ^{cl}}$ & Discontinuity of $C_V$ \\[1ex ]
\hline\hline
$0<\chi\leqslant2$& no hump over classical limit & continuous \\
$2<\chi<2.3$&no hump over classical limit & discontinuous\\
$\chi\geqslant2.3$& hump over classical limit & discontinuous\\

\hline
\end{tabular}
\end{center}
\end{table}
\\\\
In case of trapped quantum gases all  thermodynamic quantities are expressed by
polylogarithmic functions depending on fugacity 
and $ \chi$. Thus apart from fugacity, the value of $\chi$
bears the signature of difference between different quantum systems. And as seen from the theorems the value of $\chi$
dictates whether there will be a hump over the classical limit as well as the discontinuity of $C_V$.
In figure 1 we  have described the influence of different power law potentials 
on energy fluctuation of Bose system. It is clearly seen that, the ${\Delta \epsilon^2}/{kT^2 C_V ^{cl}}$
has a hump way over its classical limit when $\chi>2.3$. At $\chi=2.3$. 
the hump is just over its classical limit. There is no hump over the classical limit when $\chi<2.3$, 
All of these  are in accordance with theorem 4.1. It is also noticed that, results in  Shamyal et. al. \cite{s} are also in
agreement with the theorem. 
In their manuscript they found a hump over the classical limit in three dimensional harmonically
trapped Bose system where $\chi=\frac{3}{2}+\frac{3}{2}=3>2.3$.  
Although they did find a hump in two dimensional harmonically trapped  
Bose system but this hump was below the classical limit.
In this case, $\chi=\frac{2}{2}+\frac{2}{2}=2<2.3$ $i.e.$, no hump over the classical limit.  
Therefore, it  can  be said that the theorem 4.1 can perfectly determine whether the humps 
will be below or above the classical limit.\\\\
\begin{figure}[h!]
\centering
\includegraphics[ height=12cm, width=12cm]{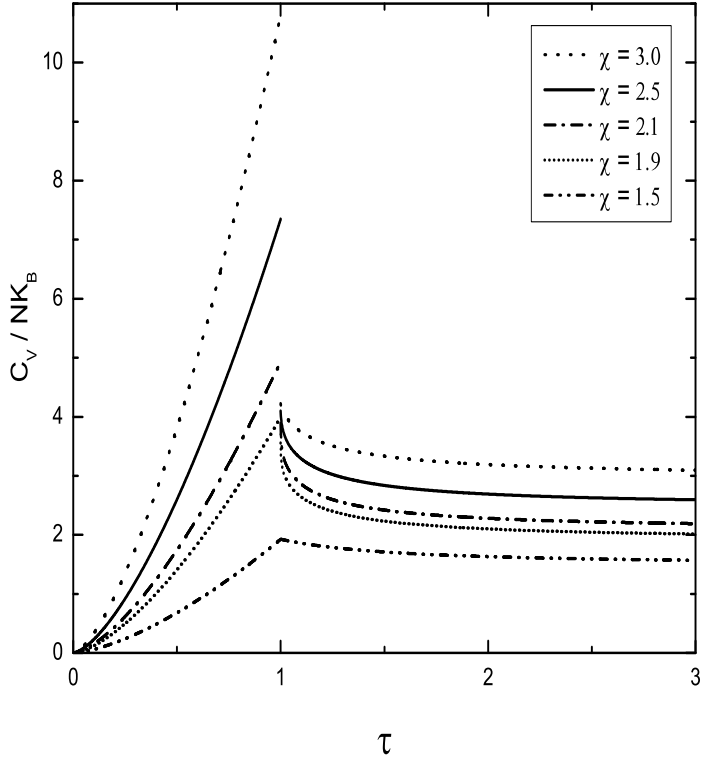}  
\caption{Specific heat of ideal trapped Bose gas
as a function of $\tau=\frac{T}{T_C}$, with different power law potentials. }
  \label{fig:boat1}
\end{figure}
Now figure 2 illustrates $C_V$ of Bose system with different trapping potentials.
It is seen from the figure that, 
$C_V$ is continuous when $\chi\leqslant2$ and it becomes discontinuous when $\chi>2$, in agreement with
theorem 4.2. 
Now, as $\chi\geqslant2.3$ denotes a hump in ${\Delta \epsilon^2}/{kT^2 C_V ^{cl}}$ over its classical limit, 
this automatically
depicts discontinuity in $C_V$. Thus, we can conclude that the appearance of a hump in
${\Delta \epsilon^2}/{kT^2 C_V ^{cl}}$
over its classical limit does indicate a
discontinuity in $C_V$ but a discontinuity in
$C_V$ does not conclude
the appearance of a hump in
${\Delta \epsilon^2}/{kT^2 C_V ^{cl}}$
over its classical limit because discontinuity in $C_V$ may arise even if $2<\chi<2.3$
but no hump in $\Delta \epsilon^2$ will exist in this interval of $\chi$. On the other hand,  $\chi\geqslant 2.3$ will      
denote a discontinuity in $C_V$ as well as 
the appearance of a hump in
${\Delta \epsilon^2}/{kT^2 C_V ^{cl}}$
over its classical limit (see table 2).\\
\section{Conclusion}
In this manuscript we have restricted our study in the case of ideal quantum gases
trapped under generic power
 law potential and proved the BMB conjecture for these types of systems.
 Point to note, as no hump in $\Delta \epsilon^2$  or no discontinuity in $C_V$ 
is noticed 
in ideal Fermi gases for any trapping potential. So, the theorems and the concluding relation 
between energy fluctuation and $C_V$ 
remain significant for ideal Bose systems only. 
It will be interesting so see the status of the above theorem for interacting quantum systems.
Also it will be very intriguing to generalize the theorems for relativistic quantum gases.
\section{Acknowledgment}
The fruitful comments of our esteemed referees are gratefully acknowledged.
MMF would also like to thank Yasmin Malik for
her cordial help to remove the minor mistakes and improve the language of the paper.

  \end{document}